\documentclass[11pt,b5paper,twoside]{article}

\usepackage[utf8]{inputenc}
\usepackage{xspace}
\usepackage{paralist}
\usepackage{algorithm}
\usepackage{algpseudocode}
\usepackage{amsthm}
\usepackage{amsmath}
\usepackage{amsfonts}
\usepackage{amssymb}
\usepackage{braket}
\usepackage[unicode]{hyperref}
\usepackage{verbatim}
\usepackage{tabularx}
\hypersetup{pdftitle=Better Algorithms for Online Bin Stretching via Computer Search}
\hypersetup{pdfauthor={Matej Lieskovský}}

\usepackage[twoside,left=2.25cm,right=1.25cm,top=1.0cm,bottom=2.0cm,papersize={182mm,257mm}]{geometry}

{\bfseries}{\itshape}

\newcommand\eps\varepsilon

\newcommand\calB{{\mathcal B}}

\newcommand\calG{{\mathcal G}}
\newcommand\calH{{\mathcal H}}
\newcommand\calI{{\mathcal I}}
\newcommand\calL{{\mathcal L}}

\newcommand\binstretch{\textsc{Online Bin Stretching}\xspace}
\newcommand\binpacking{\textsc{Bin Packing}\xspace}

\newcommand\algo{\textsc{Algorithm}\xspace}
\newcommand\adversary{\textsc{Adversary}\xspace}
\newcommand\regame{\textsc{Real Game}\xspace}
\newcommand\rogame{\textsc{Rounded Game}\xspace}
\newcommand\search{\textsc{Search}\xspace}
\newcommand\solve{\textsc{Solve}\xspace}

\newcommand\che{\textsc{Check}\xspace}
\newcommand\sort{\textsc{Sort}\xspace}
\newcommand\comp{\le_\textsc{FFD}}

\newlength{\elseskiplen}
\newcommand\elseskip{\hskip\elseskip}

\let\epsilon=\varepsilon
\def\I{\it\aftergroup\/}

\def\eps{\varepsilon}

\begin{document}
\sloppy

\setcounter{page}{1}

\title{Better Algorithms for Online Bin Stretching via Computer Search}

\author{Matej Lieskovský \thanks{\texttt{ml@iuuk.mff.cuni.cz}. Computer Science Institute of Charles University, Faculty of Mathematics and Physics, Prague, Czechia. Partially supported by GA \v{C}R project 19-27871X. }}

\date{}
\maketitle

\mbox{}
\vspace{-5ex}
\hrule height 1.5pt
\vspace{4ex}
\binstretch, introduced by Azar and Regev~\cite{AR}
is an assignment problem similar to \textsc{Online Bin Packing}.
A sequence of items of size between 0 and 1 arrive
and each must be assigned a bin before processing the next item.
We are assured that the items fit into $m$ bins of size 1.
We must also use at most $m$ bins
but we get to increase their size
to what is known as the \emph{stretching factor} $\alpha$.

Kellerer et al.~\cite{KK} showed an $\alpha = 4/3 < 1.334$ algorithm for two bins
and proved it to be optimal.
Azar and Regev~\cite{AR} showed an algorithm with $\alpha = 13/8 = 1.625$ for any $m$
and $\frac{5m-1}{3m+1}$ for $3 \le m \le 21$.
Better results have been found since then
with best known results being by Böhm et al.~\cite{BA} achieving $\alpha = 11/8 = 1.375$ for $m = 3$
and $\alpha = 3/2 = 1.5$ for all $m$.

A simple lower bound of $4/3$
was first shown for $m \in \set{2,3}$ by Kellerer et al.~\cite{KK}
and then generalised to any $m \ge 2$ by Azar and Regev~\cite{AR}.
For two bins and $\alpha < 4/3$, if the first two items are of size $1/3$,
then we either place them in the same bin and cannot place two size $2/3$ items,
or we place them in different bins and cannot place a single size 1 item.
For more bins, we can prefix the input with $m-2$ items of size 1.

Although no better lower bound for general $m$ was found,
there are results for small values of $m$.
Gabay et al.~\cite{GA} introduced the idea of using computer search
to find new lower bounds.
This approach was then improved by Böhm and Simon~\cite{BO}
who proved a lower bound of $56/41 > 1.365$ for $m=3$ and
$19/14 > 1.357$ for $4 \le m \le 8$.

We modify the computer search approach
so that it can be used for finding new algorithms for small values of $m$.
This yields new algorithms,
improving $\alpha$ for $m=4$, $m=5$ and $m=6$ significantly.

\begin{center}
\begin{tabularx}{0.909\textwidth}{ |c|c|c|c| } 
 \hline
 Number of bins & 4 & 5 & 6 \\ 
 Previous upper bound & $19/13 < 1.462$ & $3/2 = 1.5$ & $3/2 = 1.5$ \\ 
 Our upper bound & $31/22 < 1.409$ & $23/16 < 1.438$ & $19/13 < 1.462$ \\
 Granularity (see Section 1) & 22 & 16 & 13\\
 Time needed (hours) & 51 & 33 & 27 \\
 \hline
\end{tabularx}
\end{center}

The computer search was done using a server with
an Intel Xeon E5-2630 v3 CPU and 126 GB of RAM.
The time complexity grows rapidly with increasing $m$ and granularity,
but we do not believe that we have reached any fundamental limit of this method.
We are working on improving the algorithm to enable further search.

Granularity settings resulting in run times exceeding 14 days
have been tested with no success.
For three bins, we found an upper bound of $65/47 < 1.383$
using granularity 47 with a run time of 46 hours.
This is worse than the previous upper bound of $11/8 = 1.375$ by Böhm et al.~\cite{BA}
For $m=7$ we did not find an upper bound better than $3/2 = 1.5$,
mostly due to granularity being limited to at most 11.

\section{Defining \rogame}
We wish to view \binstretch as a game between two players,
\algo and \adversary.
In every round, \adversary generates an item
which \algo must place into one of the $m$ bins.
The \algo is victorious when the items generated by the \adversary
can be proven to no longer fit into $m$ offline bins of size 1
or when \adversary resigns.
The \adversary is victorious when any online bin exceeds size $\alpha$.
If a winning strategy exists for the \algo,
there exists an algorithm for \binstretch with $m$ bins
that requires a stretching factor of no more than $\alpha$.
We shall call this game the \regame.

The main obstacles to implementing a search of \regame are the following:
\begin{compactenum}
\item \adversary has an infinite selection of item sizes to pick from.
\item By sending arbitrarily small items, \adversary can make \regame last arbitrarily many rounds.
\item Proving that the items generated by \adversary so far do not fit into the offline bins can be computationally intensive.
\end{compactenum}

In order to avoid these problems,
we analyse a simplified version of \regame that we shall call \rogame
and then prove that a winning strategy for \algo in \rogame
translates to an algorithm for \binstretch.
We do this by defining \rogame in a manner that corresponds to \regame
where \algo is restricted in its decision-making process
and \adversary is permitted to cheat to a limited extent.

\rogame is parametrized by three values---the number of bins $m$,
the granularity level $k$ and the target bin size $s$---and
corresponds to \regame with $m$ bins and stretching factor $s/k$.
We scale the instance by $k$ as that permits us to work with integer values
from now on.
This makes the size of the offline bins $k$ and the size of the online bins $s$.

We classify the items by their size into $k$ classes from 0 to $k-1$,
where an item of class $c$ has size in the range $(c,c+1]$.
We also similarly define fill levels of bins from 0 to $s-1$
with the sole difference being that bins containing volume 0
are considered to have a fill level 0
while items of size 0 are disregarded entirely.
Observe that inserting a class $c$ item
will always increase the fill level of the bin by either $c$ or $c+1$.
We call the latter case an overflow.
Note that an item of class $c$ has size strictly greater than $c$
and a bin with non-zero fill level $\ell$ contains items
with total size strictly greater than $\ell$.

We would like to restrict \rogame to have a finite number of states.
Consider a state defined by the pair $(\calL,\calH)$
where $\calL$ is the list of fill levels of the individual bins
and $\calH$ is the multiset of classes of items that have been seen so far.
Let us now define the states more precisely
and show that the more precise definition indeed leads
to only a finite number of possible states.

\adversary is victorious whenever any bin's fill level exceeds $s-1$.
Since \adversary wins the game whenever a bin's fill level exceeds $s-1$,
we can restrict possible values of $\calL$
to the $s^m$ elements of $\set{0,\ldots, s-1}^m$.
This number can be further reduced
by the use of the fact that \algo is victorious
whenever the sum of all fill levels reaches $k \cdot m$,
and by observing that the order of the bins is irrelevant.

Due to the fact that there may be arbitrarily many items of class 0,
we can not include them in $\calH$.
We also know that \algo is victorious whenever the sum of item classes in $\calH$
exceeds $(k-1) \cdot m$.
These two factors alone restrict $\calH$ to being a partition
of some number smaller than $k \cdot m$,
decreasing the number of possible game states.
Note that this still includes some values of $\calH$
which correspond to items that cannot possibly fit into the offline bins.
An example of that is $m+1$ items of class $m$ with granularity $k=m+2$.
We shall deal with such histories $\calH$ later.

Having thus restricted the number of possible game states,
we now continue by defining the valid moves.
In every round of \rogame, \adversary selects an item class
and a subset of bins on which this item will cause an overflow.
\algo then selects a bin into which this item will be placed.
Note that we do not insist on the overflows being consistent
among different items and moves,
allowing \adversary to cheat in a limited manner.

We now restrict the decision-making of \algo in the following manner:
Its decisions may only depend on the current state
and the class and overflows of the item currently presented by \adversary.
Since class 0 items do not affect the state unless they overflow,
we further declare that class 0 items will be placed into
an arbitrary non-overflowing bin if such a bin exists.
We thus only have to consider the class 0 item
that causes an overflow in any bin in our analysis.
Every round now causes an increase in fill levels,
ensuring that the game is indeed finite.
This completes the definition of the \rogame.

\section{Algorithmic search}
For any given item and game state,
we use an overflow vector from $\set{0,1}^m$ to represent
which bins can the item cause to overflow.
We represent items as a pair consisting of an item's class
and an overflow item.
Since we only have to consider those class 0 items
which cause an overflow in any bin,
we precompute the set of all possible items
$\calI = \set{(0,1^m)} \cup \set{1,\ldots,k-1}\times\set{0,1}^m$.

All of our functions return a boolean value
with the convention being that $True$ indicates a victory for \algo
and $False$ indicates a victory for \adversary.
The general approach is a straightforward application
of the $\textsc{MiniMax}$ algorithm~\cite{JVN}.

Whenever a victory for \adversary is imminent,
we call the helper function $\che$,
which tries to prove that \adversary cheated
by generating a set of items that would not fit into the offline bins.
We check this by assuming that items are of the infimal size for their class
and attempting to fit them into $m$ bins of size $k-1$ by using an ILP solver.
This is computationally intensive,
which is why we do this only if \algo is about to lose.
If no such packing exists,
then the actual items cannot fit into $m$ bins of size $k$,
and we can declare a victory for \algo.

\begin{algorithm}
\caption{Automatic search for a winning strategy for \rogame}
\begin{algorithmic}[1]
\Function{\solve}{$\calL, \calH$} \Comment{Return $True$ iff game state is winnable}
\If{$\sum(\calL) \ge m \cdot k$} \Return $True$ \Comment{\adversary used too much volume}
\EndIf
\State $remaining = m \cdot k - \sum(\calL) - 1$
\If{$remaining + \min(\calL) < s$} \Return $True$ \Comment{Put all into emptiest bin}
\EndIf\label{opt-cache}
\State
\State $limit = \min(remaining, k)$
\For{$item \in \calI$}
\If{$item < limit \land \search(item, \calL, \calH) == False$}
\State \Return $False$
\EndIf
\EndFor
\State \Return $True$
\EndFunction
\State
\Function{Search}{$item, \calL, \calH$}
\For{$bin \in bins$} \Comment{Consider bins from the fullest to the emptiest}
\State $\calL' := \calL$
\State $\calL'[bin] = \calL[bin]+item.size + item.overflows[bin]$
\State $\sort(\calL')$ \Comment{Sort descending} \label{opt-sort}
\If{$\max(\calL') < t \land \solve(\calL', \calH+item.size)$} \Return $True$
\EndIf
\EndFor
\State \Return $\che(\calH)$
\EndFunction
\end{algorithmic}
\end{algorithm}

The core of our process for finding winning strategies for \algo
is the function \solve, which takes a \rogame state $(\calL,\calH)$ as input
and computes whether the game state is winnable.

\solve first tries to identify states that we already know to be won or lost.
Because a bin with a non-zero fill level $\ell$ contains
items with total size strictly greater than $\ell$,
\algo declares victory if the sum of fill levels $\calL$ is at least $m \cdot k$.
If any bin has fill level at least $k$,
it exceeded maximum permissible volume and \algo should declare a loss.
Prior to that, however, we call \che in the hope of proving that \adversary lost.
Finally, if we can identify a bin with more free space
than how much volume of items remains to be packed,
we can declare a victory by packing all remaining items into that one bin.

If \solve cannot decide the winability of the state using the processes above,
it then simulates one move of \adversary by trying every possible item.
The game state and the item is then passed to the function \search,
which simulates one move of \algo by trying every possible placement of the item
and then calling \solve recursively.

In order to attain higher granularity,
we implemented multiple optimizations.
Sorting $\calL$ in decreasing order at line \ref{opt-sort} of the algorithm
makes the successive calls of $\search$ use bins in best-fit order,
significantly improving the running time.

We also added caching of results for $\solve$.
Since caching for all possible $(\calL,\calH)$ game states used up too much memory,
we made use of the fact that, for a given game state, we can often find
either a less favourable game state that we managed to win previously
or a more favourable game state that we could not win.
We define $\calH_1 \le \calH_2$ to be true if and only if
the items from $\calH_1$ can fit into bins with sizes corresponding to the items in $\calH_2$.
Observe that this does indeed define a partial order
that (given equal fill levels) describes how good a state is for \algo.

The partial order described above is itself a generalisation of the \binpacking problem
and is thus NP-hard to evaluate.
For this reason we approximate the partial order using First-fit-decreasing~\cite{FFD}
to identify a subset of the partial order, which we shall call $\comp$.
This subset is not a partial order, but we can still use it to reduce memory and time usage.

For every given $\calL$ we store
two disjoint and initially empty sets of histories $\calG_\calL, \calB_\calL$
where $\calG_\calL$ represents won game states
and $\calB_\calL$ represents lost game states.
When evaluating the game state $(\calL, \calH)$,
we search the relevant pair of sets of histories in cache.
If there exists $\calH' \in \calG_\calL$ such that $\calH' \comp \calH$,
then the game state $(\calL, \calH')$ was winnable
and thus the better state $(\calL, \calH)$ is also winnable.
On the other hand,
if there exists $\calH' \in \calB_\calL$ such that $\calH \comp \calH'$,
then the game state $(\calL, \calH')$ was unwinnable
and thus the worse state $(\calL, \calH)$ is also unwinnable.
This cache query was used after line \ref{opt-cache} of the algorithm.
Only if the game state needed to be evaluated recursively
we add its $\calH$ to $\calG_\calL$ or $\calB_\calL$ as appropriate.

Naively caching every result improved time complexity by over 90\%
for relatively low granularity settings,
but consumed all available RAM for medium granularity settings.
LRU caching did not provide significant time complexity reductions
when compared to no caching for higher granularity settings,
probably due to the overhead soon exceeding
the rapidly decreasing probability of a cache hit.
The implementation of smart caching reduced the runtime by roughly 90\%
and memory consumption by roughly 99\%
for medium granularity settings when compared to the version with naive caching.
Smart caching consumes significantly less memory than naive caching
while not only preventing repeated evaluation of a given game state,
but also entirely preventing some game states from being evaluated
due to previously evaluated comparable game states.
This is how we achieve the results in the table.
The code is available online.~\cite{git}

\bibliographystyle{abbrv}

\end{document}